\documentclass[11pt]{article}
\usepackage[letterpaper,margin=0.76in]{geometry}
\usepackage{amsmath,amssymb,amsthm,mathtools}
\usepackage{tikz}
\usetikzlibrary{arrows.meta,calc,angles,quotes,decorations.markings,positioning}
\usepackage{microtype,booktabs,enumitem,array}
\usepackage{algorithm,algpseudocode}
\usepackage{pgfplots,placeins}
\pgfplotsset{compat=1.18}
\usepackage[colorlinks=true,linkcolor=blue,citecolor=blue,urlcolor=blue]{hyperref}
\newtheorem{theorem}{Theorem}[section]
\newtheorem{lemma}[theorem]{Lemma}
\newtheorem{proposition}[theorem]{Proposition}
\newtheorem{corollary}[theorem]{Corollary}
\theoremstyle{remark}
\newcommand{\R}{\mathbb R}
\newcommand{\CR}{\operatorname{CR}}
\newcommand{\norm}[1]{\lVert #1\rVert}
\newcommand{\Cknown}{\CR^*_{\mathrm{known}\,d}}
\newcommand{\Cunknown}{\CR^*_{\mathrm{unknown}\,d}}

\title{\bf Searching the Plane in the Segment Visibility Model
}
\author{Evangelos Kranakis\footnote{Research supported in part by Natural Sciences and Engineering Research Council of Canada.}\\
School of Computer Science, Carleton University\\
Ottawa, Ontario, K1S 5B6, Canada}
\date{}

\begin{document}
\maketitle
\begin{abstract}
We study deterministic search by $n \geq 2$ unit-speed robots starting at the origin
of the Euclidean plane. We introduce the concept of \emph{segment visibility} whereby a stationary target $x$ at distance $d=\|x\|>0$ is discovered at the first instant $t$ when $x$ belongs to the segment joining a pair of robots at time $t$. Discovery time is compared with the omniscient optimum $d$.
We distinguish two knowledge models according to whether the robots know
$d$ in advance. For $n=2$ robots, in the \emph{known-$d$} model, we give a symmetric four-phase algorithm and prove that its competitive ratio is optimal:
$\CR^*_{\mathrm{known}\,d}=1+\sqrt3+\pi/6$; the matching lower bound follows from the known-distance two-robot
shoreline-search theorem through an explicit reduction to segment detection.
For $n=2$ robots, in the \emph{unknown-$d$} model, a three-circle construction and a quantitative initial-path estimate yield the stronger lower bound $\sqrt3+\pi/6+\sqrt{17}/4
\approx3.286425990$, giving an explicit positive cost for not knowing the radius;
we also analyze a feasible antipodal-diameter logarithmic-spiral algorithm
with optimized ratio $8.928516954\ldots$, thus proving \( \sqrt3+\frac\pi6+\frac{\sqrt{17}}4\ \le\ \CR^*_{\mathrm{unknown}\,d}\ \le\ 8.928516954\ldots.\) For $n \geq 3$ robots, the competitive ratio of finding the target is exactly $\sec (\pi/n)$ regardless of whether or not $d$ is known to the robots. 

\vspace{.3cm}
\end{abstract}

\section{Introduction}\label{sec:intro}
A team of $n$ robots starts at the origin $O$ of the Euclidean plane and search for a stationary point target $x\ne O$. Each robot moves at speed at most one. The target is found precisely when at least one instantaneous closed segment joining a pair of robots contains $x$. All pairwise segments are visible in the base model.
Unlike search by one observer whose visible segment has the origin as a fixed
endpoint, both endpoints here may move independently, and the connecting
segment may pass through points never visited by either robot.

Write $d=\|x\|>0$ for the target's distance from $O$. We consider two
information regimes: in the \emph{known-$d$ model}, all robots are given
$d$ but not the direction of $x$; in the \emph{unknown-$d$ model},
neither the radius nor the direction is supplied. Both models use the
same unit-speed and visibility assumptions. Robots share a coordinate frame
and a clock, or agree on directions and a common start time before departure. For $n=2$, we show that the first regime
is exactly solvable, and that the second has optimal competitive ratio at least
$\sqrt3+\pi/6+\sqrt{17}/4$, strictly larger than the first, and at most $8.928516954\ldots$.
For unknown $d$ we analyze an antipodal logarithmic-spiral search method
whose competitive cost can be calculated in physical time. For $n \geq 3$ robots, the competitive ratio of finding the target is exactly $\sec (\pi/n)$ regardless of whether or not $d$ is known to the robots. 

\subsection{Notation and Terminology}
We introduce the necessary notation for $n=2$ robots and make the necessary adjustments for $n\geq 3$ robots in Section~\ref{app:nrobots}. Let $O=(0,0)$, and let $a,b:[0,\infty)\to\R^2$ be the positions of robots $A,B$ at \emph{physical time} $t$. The two robots start \emph{at the same time} at $O$:
\begin{equation}\label{eq:model}
a(0)=b(0)=O,\qquad
\norm{a(t)-a(s)}\le |t-s|,\quad \norm{b(t)-b(s)}\le |t-s|,
\end{equation}
for $s,t\ge0$. Thus each robot has maximal speed one. Their visibility set at time $t$ is the \emph{instantaneous} closed line segment
\(
S(t)=[a(t),b(t)]=\{(1-\lambda)a(t)+\lambda b(t):0\le\lambda\le1\}.
\)
A \emph{deterministic search algorithm} $A$ specifies a pair of robot trajectories
$a_A,b_A$ satisfying \eqref{eq:model}; before the first detection they cannot
receive any target-dependent information. Let $\mathcal A$ be the class of all
such algorithms. The letter $A$ in $\CR_A$ names an algorithm; the two robot positions of that algorithm are $a_A,b_A$.
The class $\mathcal A$ consists of strategies that do not know $d$ and includes \emph{arbitrary} admissible trajectories. After fixing an algorithm, we abbreviate its robot positions $a_A,b_A$ to $a,b$ when no confusion is possible.

For a stationary target $x\in\R^2\setminus\{O\}$, write
\(
  d=d(x):=\norm{x-O}=\norm{x}>0.
\)
Its first detection time and the two competitive ratios are defined by
\begin{equation}\label{eq:time}
\begin{aligned}
T_A(x)&:=\inf\{t\ge0: x\in[a_A(t),b_A(t)]\},~
\CR_A:=\sup_{x\ne O}\frac{T_A(x)}{d(x)},~
\CR^*:=\inf_{A\in\mathcal A}\CR_A=\CR^*_{\mathrm{unknown}\,d}.
\end{aligned}
\end{equation}
Here and everywhere below, the denominator is \emph{the offline optimum $d$}.
Since the target location
is unknown, $T_A(x)$ can be infinite for an unsuccessful algorithm.
If the exact target $x$ were given to the robots, one robot could move
straight from $O$ to $x$ in time $d$, while the other stayed at $O$.
Conversely, for any target point $x\in[a_A(t),b_A(t)]$ there is a
$\lambda\in[0,1]$ such that
\(
  d=\norm{x}
   =\norm{(1-\lambda)a_A(t)+\lambda b_A(t)}
   \le(1-\lambda)\norm{a_A(t)}+\lambda\norm{b_A(t)}
   \le t.
\)
Thus no algorithm can discover $x$ sooner
than $d$; the offline optimal detection time equals \emph{exactly $d$}.
In Section~\ref{sec:known}, an algorithm may depend on the supplied $d$;
Section~\ref{sec:unknown} considers the class $\mathcal A$ above.
In either model, the offline optimum remains exactly $d$.

If $T_A(x)<\infty$, the infimum in \eqref{eq:time} is a \emph{minimum}.
Indeed write $h_x(t):=\min_{0\le\lambda\le1}
\norm{x-(1-\lambda)a_A(t)-\lambda b_A(t)}$.
For the same $\lambda$ at times $t,s$, the two corresponding
segment points are at distance at most $|t-s|$ by the
$1$-Lipschitz motion constraints. Minimizing over $\lambda$ in
both directions shows $|h_x(t)-h_x(s)|\le|t-s|$.
Thus $h_x$ is continuous, and the detection-time set
$\{t\ge0:h_x(t)=0\}$ is closed.

Exact, scale-invariant spiral paths make infinitely many turns as $t\downarrow0$. They are nonetheless well-defined, continuous, $1$-Lipschitz trajectories starting at $O$ at $t=0$. If a finite initial number of maneuvers is required, one may instead impose a known positive lower bound on $d$ and modify the initialization; exact multiplicative ratios then require a separate initialization analysis.

The algorithms below are \emph{continuous-time trajectory specifications} in that for every physical time $t>0$ the displayed formula prescribes each robot's location, and a continuous sensor reports the first instant at which the unknown target belongs to the indicated segment.  We denote this sensor by
$\operatorname{Hit}([P,Q])$; the sensor returns true exactly if the unknown target lies in $[P,Q]$, without revealing the target's position. In every algorithm the robots start at $O$.
Algorithm~\ref{alg:known_optimal} uses the specified distance $d$ as input;
Algorithm~\ref{alg:antipodal} does not.
The spiral expressions involving $\log t$ apply only for $t>0$.

\subsection{Literature Review}\label{subsec:related}

Competitive geometric search has its origins in search games, including
Gal's minimax framework for linear and planar searching~\cite{alperngal,baezaculber,gal1974,gal1980}. Note that single robot search in the plane can be thought of as segment visibility whereby one of the searchers stays fixed at the origin. 
For such a single unit-speed searcher with origin-to-observer radial visibility,
logarithmic-spiral search is known to have an optimal ratio of approximately
$17.28935\ldots$. Gal's work supplies the classical spiral construction,
whereas Langetepe~\cite{langetepe2010} proves optimality among unrestricted
single-observer strategies. That visibility rule is different from the
present problem: our observed segment joins \emph{two moving robots},
and a target may be discovered at an interior chord point.

A particularly useful comparison is \emph{shoreline search}, in which
mobile robots attempt to touch an unknown line in the plane and the
performance denominator is the line's distance from the origin~\cite{finch,isbell}.
For two robots Acharjee, Georgiou, Kundu, and Srinivasan~\cite{acharjee2020}
proved a universal lower bound of $3$.
The subsequent work of Dobrev, Kr\'alovi\v c, and
Pardubsk\'a~\cite{dobrev2020} sharpened the two-robot lower bound to
$1+\sqrt3+\pi/6$ when the shoreline distance is known, matching the
corresponding shoreline upper bound. Sections~\ref{sec:known} and~\ref{app:nrobots} explain how these shoreline results yield lower bounds for segment search. 
In Section~\ref{sec:known}, we make use of that theorem to prove a lower bound for our problem by reduction to the shoreline problem. A potentially useful method can be found in the recent preprint of Temerev~\cite{tamerev2026}; its main theorem shows that for every $d_0 > 0$, every deterministic path which searches all straight lines at distances $d \geq d_0$ has competitive ratio at least $12.5937096\ldots$, but its approach does not directly improve our two-robot lower bound.  

Multi-agent spiral methods, including variants with different agent speeds, have also been studied~\cite{georgiou2026}. The detection rules in this literature must be compared explicitly; trajectories designed for physical point encounters or for fixed-origin visibility cannot be
transferred automatically to moving-chord visibility. 

In the unknown-radius
model we study a pair of antipodal spirals, for which the joining chord is
a full diameter. This family is not claimed optimal among arbitrary
two-robot trajectories when $d$ is unknown.

\subsection{Results and Organization}\label{subsec:results}
For two robots, we first determine the exact deterministic optimum for a known target
radius, $1+\sqrt3+\pi/6\approx3.255649583$. For unknown $d$, we prove the stronger unconditional lower bound
$\sqrt3+\pi/6+\sqrt{17}/4\approx3.286425990$. The resulting separation
from the known-radius optimum is at least $(\sqrt{17}-4)/4\approx0.030776406$.
The best upper bound developed here for the latter model is achieved by
antipodal logarithmic spirals. 
\begin{center}
\small
\begin{tabular}{@{}p{.29\textwidth}p{.53\textwidth}@{}}\toprule
\textbf{Model} & \textbf{Established Competitive Ratio}\\\midrule
$n=2$: Known $d$ 
& $\displaystyle\CR^*_{\mathrm{known}\,d}=1+\sqrt3+\frac\pi6$ \\[2pt]
$n=2$: Unknown $d$ &
$\displaystyle \frac{\sqrt{17}}4 +\sqrt3+\frac\pi6 \le\CR^*_{\mathrm{unknown}\,d}\le8.928516954\ldots$\\
$n\geq 3$: Known/Unknown $d$ & $\sec(\pi/n)$\\
\bottomrule
\end{tabular}
\end{center}
Section~\ref{sec:known} proves the known-radius tight result, with
algorithmic pseudocode, feasibility, coverage, and the shoreline reduction.
Section~\ref{sec:unknown} analyzes the unknown-radius antipodal spiral
algorithm and proves the explicit three-circle lower bound. Section~\ref{app:nrobots} extends the results to $n \geq 3$ searchers.
Section~\ref{sec:conclusion} summarizes the results and proposes some open questions. 

\section{Two Robots: Known Distance of Target from the Origin}\label{sec:known}

Here we analyze the competitive ratio of ray search when both robots are
informed of the target radius $d>0$ but not its angular position. The
value $d$ may be used when choosing the trajectory pair; the target is
discovered when it belongs to their current joining segment. We establish
matching upper and lower bounds for \emph{all} deterministic admissible
strategies, not just the symmetric algorithm used for the upper bound.

Let $\mathfrak A_d$ denote the class of admissible algorithms allowed to
use the known parameter $d$. For the circle
$S_d=\{p\in\mathbb R^2:\|p\|=d\}$ and each
$A\in\mathfrak A_d$, define
\begin{align*}
T_{A}(p)&:=\inf\{t\ge0:p\in[a(t),b(t)]\},~
\CR_{A,d}:=\sup_{p\in S_d}\frac{T_{A}(p)}d,~
\CR^*_{\mathrm{known}\,d}:=\inf_{A\in\mathfrak A_d}\CR_{A,d}.
\end{align*}
The offline optimum for each $p\in S_d$ is $d$ (Section~\ref{sec:intro}).
For convenience set $C_0=1+\sqrt3+\pi/6$.

\begin{theorem}[Optimal deterministic search with known target radius]
\label{thm:known_chord}\label{thm:known_tight}
For every specified $d>0$,
\begin{equation}\label{eq:known_radius_bounds}
 \CR^*_{\mathrm{known}\,d}=1+\sqrt3+\frac\pi6.
\end{equation}
Algorithm~\ref{alg:known_optimal} attains this ratio, with exact worst-case
discovery time $C_0d$. The matching lower bound holds for all deterministic
admissible trajectories, without symmetry, monotonicity, or boundary-following
assumptions.
\end{theorem}

We prove the upper bound in Subsection~\ref{subsec:known_upper}. 
For the lower bound in Subsection~\ref{subsec:known_lower}, we invoke the
known-distance shoreline theorem of Dobrev, Kr\'alovi\v c, and
Pardubsk\'a~\cite{dobrev2020}, and give a detailed reduction from segment
discovery to shoreline search. 

\subsection{Upper Bound}\label{subsec:known_upper}

Define the waypoints
\[
 A_\pm=\left(\pm\frac d{\sqrt3},d\right),
 B_\pm=\left(\pm\frac{\sqrt3d}{2},\frac d2\right),
 C_\pm=(\pm d,0), 
 D_\pm=(\pm d,-d).
\]
Each robot follows the corresponding trajectory
straight segment $OA_\pm$, the tangent
segment $A_\pm B_\pm$, the short circular arc $B_\pm C_\pm$, and finally
the vertical segment $C_\pm D_\pm$. Both robots execute the phases
simultaneously at unit speed and the operator $\operatorname{Hit}([a(t),b(t)])$ detects whether or not the target lies in the segment $[a(t),b(t)]$. 

\begin{algorithm}[H]
\caption{Optimal segment search for a target at known radius $d$}
\label{alg:known_optimal}
\begin{algorithmic}[1]
\Require Known target radius $d>0$; continuous segment-detection sensor.
\State Start both robots at $O$ and assign signs $\sigma\in\{-1,+1\}$.
\State Throughout all phases, monitor $\operatorname{Hit}([a(t),b(t)])$;
       on the first hit, report discovery and stop both robots.
\State In parallel, robot $\sigma$ travels straight to
       $A_\sigma=(\sigma d/\sqrt3,d)$ at unit speed.
\State In parallel, robot $\sigma$ travels straight to
       $B_\sigma=(\sigma\sqrt3d/2,d/2)$ at unit speed.
\State In parallel, robot $\sigma$ follows the circle towards
       $C_\sigma=(\sigma d,0)$ at unit speed.
\State In parallel, robot $\sigma$ travels vertically to
       $D_\sigma=(\sigma d,-d)$ at unit speed.
\State Every possible target has been detected by this time.
\end{algorithmic}
\end{algorithm}

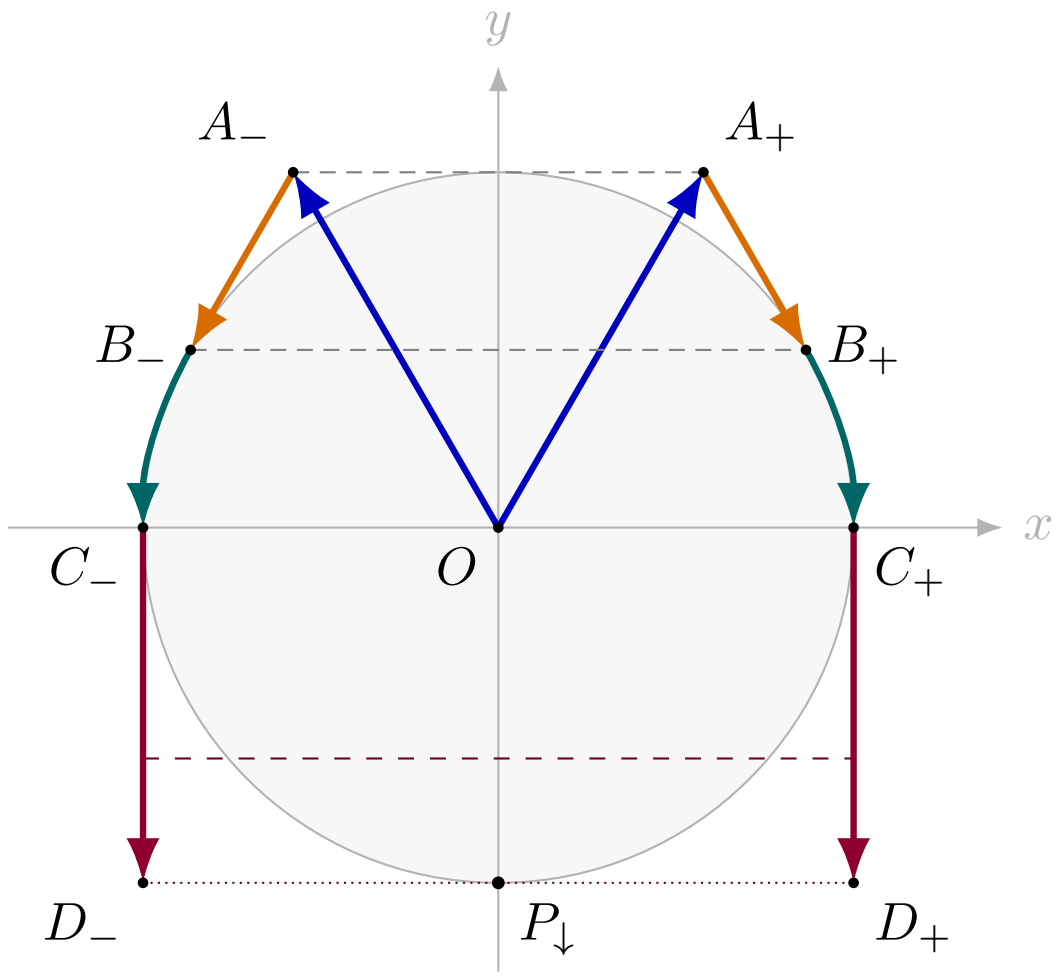
\begin{figure}[htbp]
\centering
\begin{tikzpicture}[scale=2.35,>=Latex,font=\small]
 \fill[gray!6] (0,0) circle (1);
 \draw[gray!60] (0,0) circle (1);
 \draw[->,gray!60] (-1.38,0)--(1.42,0) node[right] {$x$};
 \draw[->,gray!60] (0,-1.25)--(0,1.3) node[above] {$y$};
 \draw[blue!75!black,very thick,->] (0,0)--(.577350,1);
 \draw[blue!75!black,very thick,->] (0,0)--(-.577350,1);
 \draw[orange!85!black,very thick,->] (.577350,1)--(.866025,.5);
 \draw[orange!85!black,very thick,->] (-.577350,1)--(-.866025,.5);
 \draw[teal!80!black,very thick,->] (.866025,.5) arc (30:0:1);
 \draw[teal!80!black,very thick,->] (-.866025,.5) arc (150:180:1);
 \draw[purple!75!black,very thick,->] (1,0)--(1,-1);
 \draw[purple!75!black,very thick,->] (-1,0)--(-1,-1);
 \draw[densely dashed,gray] (-.577350,1)--(.577350,1);
 \draw[densely dashed,gray] (-.866025,.5)--(.866025,.5);
 \draw[purple!60!black,dashed] (-1,-.65)--(1,-.65);
 \draw[purple!60!black,densely dotted] (-1,-1)--(1,-1);
 \fill (0,0) circle (.015) node[below left] {$O$};
 \fill (.577350,1) circle (.015) node[above right] {$A_+$};
 \fill (-.577350,1) circle (.015) node[above left] {$A_-$};
 \fill (.866025,.5) circle (.015) node[right] {$B_+$};
 \fill (-.866025,.5) circle (.015) node[left] {$B_-$};
 \fill (1,0) circle (.015) node[below right] {$C_+$};
 \fill (-1,0) circle (.015) node[below left] {$C_-$};
 \fill (1,-1) circle (.015) node[below right] {$D_+$};
 \fill (-1,-1) circle (.015) node[below left] {$D_-$};
 \fill (0,-1) circle (.018) node[below right] {$P_\downarrow$};
\end{tikzpicture}
\caption{Optimal known-radius discovery, drawn for $d=1$. After straight
deployment (blue) and tangent motion (orange), the robots follow only the
short arcs to the equator (teal), then move vertically downwards (purple).
The dashed horizontal segment illustrates the final scan of the lower
semicircle. The bottom point $P_\downarrow=(0,-d)$ is the worst-case target.}
\label{fig:known_optimal}
\end{figure}

Figure~\ref{fig:known_optimal} illustrates
the complete construction.
\FloatBarrier

\subsubsection{Physical-time trajectories and feasibility}

Define the transition times
\[
 t_1=\frac{2d}{\sqrt3},~
 t_2=\sqrt3d,~
 t_3=\left(\sqrt3+\frac\pi6\right)d,~
 t_4=C_0d.
\]
For $\sigma\in\{-1,+1\}$, the prescribed position of robot $\sigma$ at time $t$ is given by
\begin{equation}\label{eq:known_chord_trajectories}
 p_\sigma(t)=
 \begin{cases}
  (\sigma t/2,\sqrt3t/2),&0\le t\le t_1,\\
  \left(\sigma\left(d/\sqrt3+(t-t_1)/2\right),
       d-\sqrt3(t-t_1)/2\right),&t_1\le t\le t_2,\\
  (\sigma d\sin\theta(t),d\cos\theta(t) ),
       \theta(t)=\pi/3+(t-t_2)/d,&t_2\le t\le t_3,\\
  (\sigma d,-(t-t_3)),&t_3\le t\le t_4.
 \end{cases}
\end{equation}
Set $a(t)=p_+(t)$ and $b(t)=p_-(t)$. After $t_4$, keep the prescribed
no-detection positions fixed. These formulas are interrupted at discovery.

\begin{lemma}[Admissibility]\label{lem:known_feasible}
The trajectories in~\eqref{eq:known_chord_trajectories} start at $O$
simultaneously, are continuous, and $1$-Lipschitz.
\end{lemma}
\begin{proof}
The formulas give $O$ at $t=0$ and agree at the successive waypoints
$A_\sigma$, $B_\sigma$, $C_\sigma$, and $D_\sigma$.
The velocity norms in the first two phases are respectively
$\sqrt{(1/2)^2+(\sqrt3/2)^2}=1$ and the same quantity.
In the circular phase, $\theta'(t)=1/d$, so
\(
 p'_\sigma(t)=(\sigma\cos\theta(t),-\sin\theta(t))
 \)
 and
\( \norm{p'_\sigma(t)}=1.
\)
The final velocity is $(0,-1)$. Therefore each path is continuous and
piecewise continuously differentiable with speed at most one. Integration
over any time interval, splitting at transition times if necessary, gives
$\norm{p_\sigma(t)-p_\sigma(s)}\le|t-s|$.
\end{proof}

\subsubsection{Coverage and exact worst-case discovery time}

\begin{proposition}\label{prop:known_upper}
Algorithm~\ref{alg:known_optimal} discovers every point of the circle $S_d$ by time $C_0d$, and its exact worst-case discovery time is $C_0d$, where $C_0=1+\sqrt3+\pi/6$.
\end{proposition}
\begin{proof}
We verify coverage of three portions of $S_d$ in order. Any detection
during the initial deployment only improves the search time.

\emph{Upper cap, $d/2\le y\le d$.}
During the tangent phase, put $u=t-t_1\in[0,d/\sqrt3]$. The endpoints are
$(\pm X(u),Y(u))$, where
\[
 X(u)=\frac d{\sqrt3}+\frac u2,\qquad
 Y(u)=d-\frac{\sqrt3u}{2}.
\]
Their squared distance from the origin satisfies
\[
 X(u)^2+Y(u)^2=d^2+\left(u-\frac d{\sqrt3}\right)^2\ge d^2.
\]
Since $X(u)\ge0$, this implies
$X(u)\ge\sqrt{d^2-Y(u)^2}$. Thus the horizontal connecting segment
contains both circle points $(\pm\sqrt{d^2-Y(u)^2},Y(u))$.
The height $Y(u)$ decreases continuously from $d$ to $d/2$, covering
the entire upper cap by $t_2$.

\emph{Remaining upper semicircle, $0\le y\le d/2$.}
During the circular phase $\theta$ increases from $\pi/3$ to $\pi/2$.
The endpoints themselves visit every point of the two corresponding arcs.
An endpoint is part of the closed visibility segment. Consequently, by $t_3$
the entire upper semicircle has been discovered.

\emph{Lower semicircle, $-d\le y\le0$.}
During the final phase write $s=t-t_3\in[0,d]$. The visibility segment is
\begin{equation}\label{eq:horizontal_scan}
 [a(t),b(t)]=\{(z,-s):-d\le z\le d\}.
\end{equation}
It contains both points $(\pm\sqrt{d^2-s^2},-s)$ on the target circle.
As $s$ runs from zero to $d$, this covers the lower semicircle. The coverage
argument, together with Lemma~\ref{lem:known_feasible}, proves
$\sup_{p\in S_d}T_A(p)\le t_4=C_0d$ for Algorithm~\ref{alg:known_optimal}.

For equality, place the target at $P_\downarrow=(0,-d)$. In the first three
phases both endpoints have nonnegative height, and so does every point of
their segment. In the last phase the segment has height $-s>-d$ whenever
$s<d$. Therefore $P_\downarrow$ is not detected before $t_4$ and is detected
at $t_4$. This proves the exact worst-case assertion.
\end{proof}

\subsection{Lower Bound}\label{subsec:known_lower}

The shoreline problem has the same start and speed constraints, but the
``hidden object'' is an infinite line at known distance $d$ from $O$.
The line is found when either robot physically touches it. Observe that
every such line
can be written uniquely as
\(
 \ell_{u,d}=\{z\in\mathbb R^2:u\cdot z=d\},~ \norm u=1,
\)
where $u$ points from $O$ towards the nearest point on the line. Indeed, the line passes through \(p=du\). For any other point \(z\) on it, the displacement \(z-p\) lies along the line and is therefore perpendicular to \(u\), i.e.,
\(
u\cdot(z-p)=0.
\)
Substituting \(p=du\) gives
\(
u\cdot z=d(u\cdot u)=d,
\)
because \(\|u\|=1\).

Fix a deterministic search algorithm, and let \(a,b\) be the trajectories it prescribes under the hypothetical assumption that no target is ever detected. Define the shoreline hitting time
\begin{equation}
\label{eq:hittingtime}
 H_{a,b}(u;d)=\inf\{t\ge0:a(t)\in\ell_{u,d}
                         \text{ or }b(t)\in\ell_{u,d}\}.
\end{equation}

In the sequel, we make use of the following two-agent known-distance lower bound proved by Dobrev, Kr\'alovi\v c, and Pardubsk\'a~\cite{dobrev2020}. 
\begin{theorem}[\cite{dobrev2020}]
\label{thm:known_shoreline}
For any two admissible deterministic robot trajectories starting at $O$,
even when they are chosen with knowledge of $d$,
\[
 \sup_{\norm u=1}\frac{H_{a,b}(u;d)}d
 \ge1+\sqrt3+\frac\pi6.
\]
\end{theorem}

We invoke their theorem for unrestricted trajectories; in particular, we do not assume that each
robot searches a prescribed semicircle. The following reduction is proved below.

For fixed paths $a,b$, write $T_{a,b}(x):=\inf\{t\ge0:x\in[a(t),b(t)]\}$.
If these are the paths of algorithm $A$, then $T_{a,b}(x)=T_A(x)$.

\begin{figure}[htbp]
\centering
\begin{tikzpicture}[scale=1.9,>=Latex,font=\small]
 \fill[gray!6] (0,0) circle (1);
 \draw[gray!60] (0,0) circle (1);
 \draw[gray!60,->] (-1.18,0)--(1.95,0);
 \draw[red!75!black,thick] (1,-1.3)--(1,1.3)
     node[above] {$\ell_{u,d}:u\cdot z=d$};
 \draw[blue!75!black,very thick] (.25,-.6)--(1.75,.6);
 \draw[->,thick] (0,0)--(.55,0) node[midway,above] {$u$};
 \draw[gray,dashed] (0,0)--(1,0);
 \fill (0,0) circle (.018) node[below left] {$O$};
 \fill (.25,-.6) circle (.023) node[below left] {$b(T)$};
 \fill (1.75,.6) circle (.023) node[right] {$a(T)$};
 \fill[red!75!black] (1,0) circle (.025) node[below right] {$p=du$};
\end{tikzpicture}
\caption{Reduction to shoreline search, shown with $d=1$ and $u=(1,0)$.
If the current segment contains $p=du$, at least one endpoint has projection
at least $d$ along $u$. Since that endpoint started at $O$, it must have
touched the tangent line $\ell_{u,d}$ at or before this detection time.
The displayed configuration is illustrative, not part of the upper-bound
algorithm.}
\label{fig:known_reduction}
\end{figure}
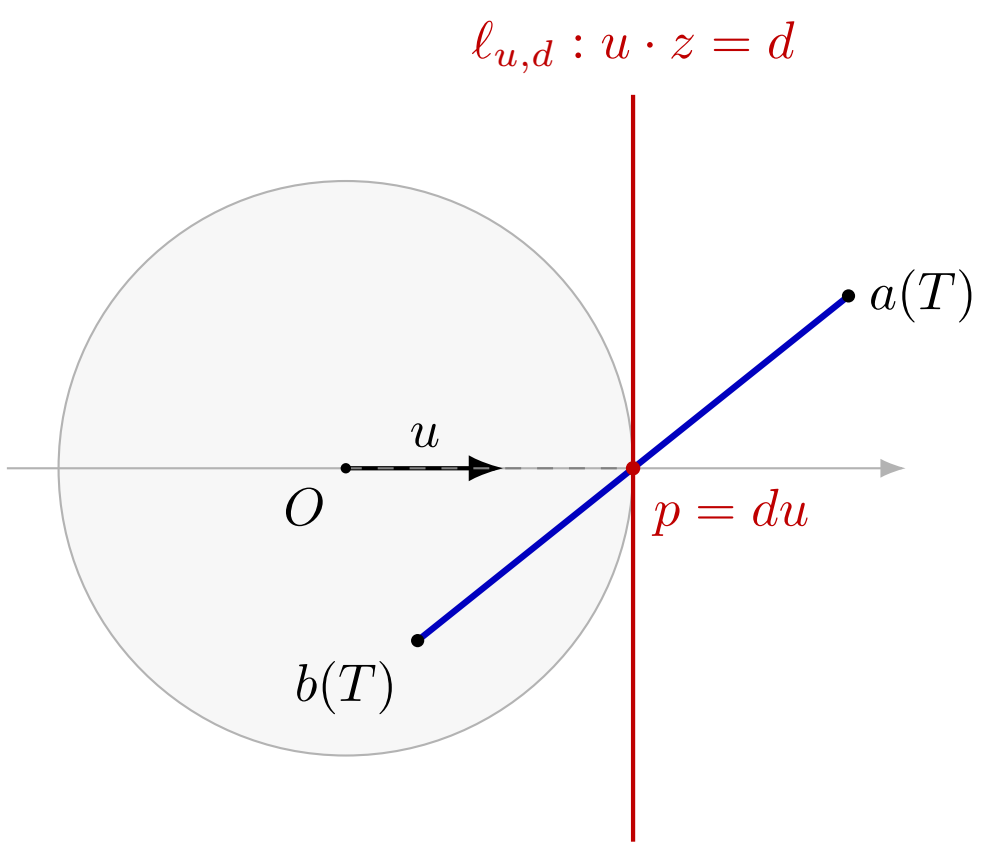

\begin{lemma}[A tangent line is reached no later than its tangency target]
\label{lem:known_reduction}
For every admissible trajectory pair $a,b$, every $d>0$,
and every unit vector $u\in\mathbb{R}^2$,
\begin{equation}\label{eq:known_reduction}
H_{a,b}(u;d)\le T_{a,b}(du).
\end{equation}
\end{lemma}
\begin{proof}
Let $A$ denote the algorithm generating $a,b$.

If $T_{A}(du)=\infty$, then the inequality holds. Otherwise put $T=T_{A}(du)$. Continuity of the moving
segment ensures that the first detection is attained, as justified in
Section~\ref{sec:intro}. Hence, for some $\lambda\in[0,1]$,
\(
 du=(1-\lambda)a(T)+\lambda b(T).
\)
Taking the scalar product with $u$ and using $\norm u=1$ yields
\(
 d=(1-\lambda)u\cdot a(T)+\lambda u\cdot b(T)
 \le\max\{u\cdot a(T),u\cdot b(T)\}.
\)
In the sequel, refer to Figure~\ref{fig:known_reduction}.
Choose $g\in\{a,b\}$ such that its projection at $T$ is at
least $d$. Its scalar projection $s\mapsto u\cdot g(s)$ is continuous and
starts at zero. Since $d>0$, the intermediate value theorem gives
$s\in[0,T]$ with $u\cdot g(s)=d$. Thus $g(s)\in\ell_{u,d}$ and
$H_{a,b}(u;d)\le s\le T$, proving the claim.
\end{proof}

\begin{proof}[Proof of Theorem~\ref{thm:known_tight}]
Fix any deterministic segment-search algorithm $A$ that knows
the target radius $d$. Let $a,b$ be the trajectories prescribed
by $A$ under the hypothetical assumption that no target is ever
detected. These trajectories are independent of the target's
unknown direction.

Construct a shoreline-search algorithm by having the robots
follow $a,b$ until one of them reaches the hidden line.
This is a valid known-distance shoreline-search algorithm:
its motion schedule uses the given distance $d$, but does not
use the unknown location or orientation of the line.

Lemma~\ref{lem:known_reduction} holds for every direction $u$, so
\begin{align*}
 \CR_{A,d}
&=\sup_{\norm u=1}\frac{T_{A}(du)}d
\ge\sup_{\norm u=1}\frac{H_{a,b}(u;d)}d
\ge1+\sqrt3+\frac\pi6,
\end{align*}
where the last inequality is Theorem~\ref{thm:known_shoreline}. Taking the
infimum over all $A$ gives the desired universal lower bound.
Proposition~\ref{prop:known_upper} gives an admissible algorithm attaining
the same value, which proves equality.
\end{proof}

\section{Two Robots: Unknown Distance of Target from the Origin}\label{sec:unknown}

We now require the robots to search without knowing either the target's
distance $d$ or its angular direction. 

\subsection{Upper Bound}

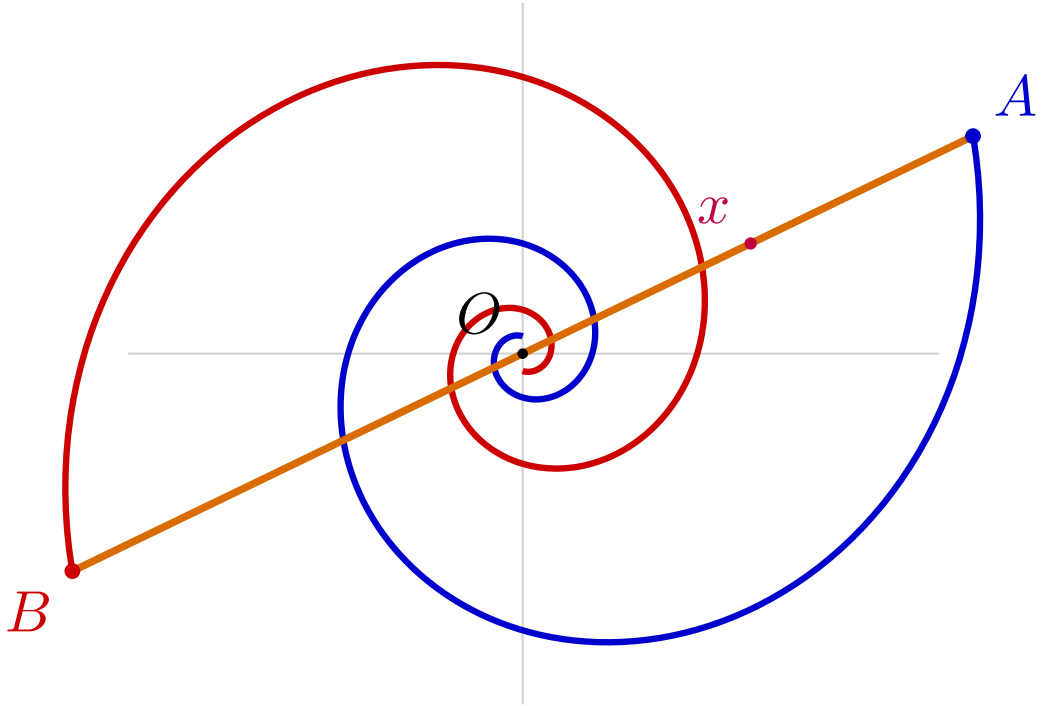
\begin{figure}[htbp]\centering
\begin{tikzpicture}[>=Latex]
\begin{scope}[scale=2.9]
\draw[gray!35] (-.9,0)--(.95,0) (0,-.8)--(0,.8);
\draw[blue!80!black,very thick,domain=-11:.45,samples=300,variable=\s]
plot ({exp(.293124*\s)*cos(\s r)},{exp(.293124*\s)*sin(\s r)});
\draw[red!80!black,very thick,domain=-11:.45,samples=300,variable=\s]
plot ({-exp(.293124*\s)*cos(\s r)},{-exp(.293124*\s)*sin(\s r)});
\coordinate (O) at (0,0);
\coordinate (A) at (1.027,.496);
\coordinate (B) at (-1.027,-.496);
\draw[orange!85!black,line width=1.4pt] (A)--(B);
\fill (O) circle (.012) node[above left] {$O$};
\fill[blue!80!black] (A) circle (.018) node[above right] {$A$};
\fill[red!80!black] (B) circle (.018) node[below left] {$B$};
\fill[purple] (.52,.2513) circle (.014) node[above left] {$x$};
\end{scope}
\end{tikzpicture}
\caption{Two \emph{different} rotated logarithmic spirals in Algorithm~\ref{alg:antipodal}, one followed by each robot at speed one. The chord contains the origin $O$ at every time and detects any target on the current diameter. }
\label{fig:antipodal}
\end{figure}

Unlike the preceding four-phase
algorithm, a valid strategy here must prescribe its trajectories without
using $d$. We study a scale-invariant logarithmic-spiral construction:
the robots follow two antipodal spirals with the visibility segment
passing through $O$ (see Figure~\ref{fig:antipodal}). 

The construction uses physical time and works for every $d>0$.
Choose $\lambda>0$, a fixed reference time $t_*>0$, and a phase
$\theta_*\in\R$. These choices are independent of the target. Put
\begin{equation}\label{eq:spiralphysical}
v=\frac{\lambda}{\sqrt{1+\lambda^2}},\qquad
\theta(t)=\theta_*+\frac1\lambda\log(t/t_*),\qquad r(t)=vt\quad(t>0).
\end{equation}
The prescribed paths are
\begin{equation}\label{eq:spiralpaths}
a(0)=b(0)=O,\qquad
a(t)=r(t)(\cos\theta(t),\sin\theta(t)),\qquad b(t)=-a(t)\quad(t>0).
\end{equation}
The reference time merely fixes a normalization; one may take $t_*=1$
and $\theta_*=0$. Eliminating time gives the logarithmic spiral
$r=vt_*e^{\lambda(\theta-\theta_*)}$. The angle has no finite limit at
zero, but the positions converge to $O$. No positive initial radius or
limiting initialization is used.

\begin{algorithm}[H]
\caption{Antipodal logarithmic-spiral search}
\label{alg:antipodal}
\begin{algorithmic}[1]
\Require $\lambda>0$; fixed $t_*>0$ and $\theta_*\in\R$; continuous segment-detection sensor
\State Set $v\gets\lambda/\sqrt{1+\lambda^2}$
\State Start both robots at $a(0)=b(0)=O$ at time $0$
\While{no target has been detected}
\State For physical time $t>0$, set $r(t)=vt$ and $\theta(t)=\theta_*+\lambda^{-1}\log(t/t_*)$
\State Move simultaneously along $a(t)=r(t)(\cos\theta(t),\sin\theta(t))$ and $b(t)=-a(t)$
\State Continuously monitor $\operatorname{Hit}([a(t),b(t)])$
\EndWhile
\State Announce detection and stop both robots
\end{algorithmic}
\end{algorithm}

The word \emph{antipodal} means that the robots occupy opposite ends of
a diameter at every positive physical time. The instantaneous visibility
segment $[-a(t),a(t)]$ is the union of two radial segments. Robot $B$
does not traverse the same geometric spiral as robot $A$: its track is
the half-turn rotation of $A$'s track.

\begin{theorem}\label{thm:antipodal}
For every $\lambda>0$, Algorithm~\ref{alg:antipodal} is admissible in the
start-at-origin model and has exact competitive ratio
\begin{equation}\label{eq:spiralratio}
f(\lambda)=\frac{e^{\pi\lambda}\sqrt{1+\lambda^2}}{\lambda}.
\end{equation}
Its unique minimum over this family occurs at the positive solution
$\lambda^*$ of $\pi\lambda^*(1+(\lambda^*)^2)=1$, where
$\lambda^*\approx0.293124142982$ and
$f(\lambda^*)\approx8.928516954107$.
\end{theorem}
\begin{proof}
For a logarithmic spiral, radial distance grows exponentially with the angle: $r = a e^{\lambda \theta}$. Hence, $\frac{dr}{d\theta} = \lambda r$. The arc length in polar coordinates is
\(
   \frac{ds}{d\theta}
   =\sqrt{r^2+\left(\frac{dr}{d\theta}\right)^2}
   =r\sqrt{1+\lambda^2}.
\)
Since the robot moves at unit speed,
\(
   1=\frac{ds}{dt}
   =r\sqrt{1+\lambda^2}\,\dot\theta,
\)
hence
\(
   \dot\theta=\frac{1}{r\sqrt{1+\lambda^2}}
\)   
and
\(   
   \dot r=\frac{dr}{d t} =\frac{dr}{d\theta}\dot\theta
       =\lambda r\cdot \frac{1}{r\sqrt{1+\lambda^2}}
       =\frac{\lambda}{\sqrt{1+\lambda^2}}.
\)
Thus the radial speed (i.e., the rate at which its distance from the origin increases) is constant:
\(
   v=\frac{\lambda}{\sqrt{1+\lambda^2}}.
\)

The two robots are antipodal, so the detecting segment is the diameter
\([a(t),b(t)]\), whose supporting line is the ray of angle \(\theta(t)\)
modulo \(\pi\). Therefore a target at polar coordinates \((d,\phi)\) is
detected exactly when
\(
   \theta(t)\equiv \phi \pmod{\pi}
   \quad\text{and}\quad
   d\le r(t).
\)
For a fixed target direction \(\phi\), the successive detection angles are
\(
   \theta_k=\phi+k\pi,
   \qquad k\in\mathbb Z.
\)
The corresponding radii are
\[
   R_k=r_0e^{\lambda(\phi-\theta_0+k\pi)}
      =R_0q^k,
   \qquad
   q=e^{\lambda\pi}, \qquad R_0 = e^{\lambda(\phi-\theta_0)}.
\]
Thus consecutive possible detection radii differ by the factor
\(
   \frac{R_{k+1}}{R_k}=e^{\lambda\pi}=q.
\)

Let \(d\) be the unknown target distance, and let \(k\) be the first index
with \(R_k\ge d\). Then the algorithm detects the target at time
\(
   t_k=\frac{R_k-r_0}{v}.
\)
Since \(R_{k-1}<d\), we have
\(
   R_k=qR_{k-1}<qd.
\)
Therefore
\(
   t_k
   \le \frac{qd-r_0}{v}
   =\frac{q}{v}\,d-\frac{r_0}{v}.
\)
Ignoring the additive initial-radius term, or taking \(r_0\to0\), the
algorithm's time is at most
\(
   \frac{q}{v}\,d.
\)
The offline optimum is at least \(d\), because the target is at distance
\(d\) from the origin and the searcher has unit speed. Hence
\[
   \operatorname{CR}(\lambda)
   \le  \frac{qd/v}d = \frac{q}{v}
   =\frac{e^{\lambda\pi}\sqrt{1+\lambda^2}}{\lambda}.
\]

Conversely, fix \(\varepsilon>0\) and choose a target distance
\(
   d=(1+\varepsilon)R_{k-1}.
\)
Then the target is not detected at radius \(R_{k-1}\), and it is first
detected at radius
\(
   R_k=qR_{k-1}
      =\frac{q}{1+\varepsilon}\,d.
\)
The algorithm's time divided by the offline time is therefore
\[
   \frac{R_k/v}{d}
   =\frac{q}{v(1+\varepsilon)}
   \longrightarrow \frac{q}{v}
   \qquad(\varepsilon\to0).
\]
Thus, the competitive ratio can be expressed as 
\(
   \operatorname{CR}(\lambda)
   =\frac{e^{\lambda\pi}\sqrt{1+\lambda^2}}{\lambda}.
\)

It remains to choose \(\lambda>0\) optimally. Define
\(
   f(\lambda)
   :=\frac{e^{\lambda\pi}\sqrt{1+\lambda^2}}{\lambda}.
\)
Then
\(
   \ln f(\lambda)
   =\lambda\pi+\frac12\ln(1+\lambda^2)-\ln\lambda,
\)
so
\(
   \frac{d}{d\lambda}\ln f(\lambda)
   =\pi+\frac{\lambda}{1+\lambda^2}-\frac{1}{\lambda}.
\)
Setting this equal to zero gives
\(
   \pi+\frac{\lambda}{1+\lambda^2}-\frac{1}{\lambda}=0,
\)
or equivalently
\(
   \pi\lambda(1+\lambda^2)=1.
\)
The unique positive solution is
\(
   \lambda^\star\approx 0.293124142982\ldots.
\)
At this value, the competitive ratio is
\[
   \frac{e^{\lambda^\star\pi}\sqrt{1+(\lambda^\star)^2}}{\lambda^\star}
   \approx 8.928516954107\ldots,
\]
which completes the proof of Theorem~\ref{thm:antipodal}.

\end{proof}

\subsection{Lower Bound}\label{subsec:unknown_lower}

Recall that the two robots start simultaneously at $O=(0,0)$ and follow paths $a,b:[0,\infty)\to\R^2$ satisfying
\begin{equation}\label{gap:eq:admissible}
a(0)=b(0)=O,\qquad
\norm{a(t)-a(s)}\le |t-s|,\quad
\norm{b(t)-b(s)}\le |t-s|\qquad(s,t\ge0);
\end{equation}
they have maximum speed of one and their paths are $1$-Lipschitz. For a target $x\ne O$, define
\begin{equation}\label{gap:eq:detection}
T_{a,b}(x):=\inf\{t\ge0:x\in[a(t),b(t)]\},\qquad
\inf\varnothing:=+\infty.
\end{equation}
The offline discovery time is $\norm x$: one robot can go straight to $x$, and no connecting segment can contain a point farther than $t$ from $O$ at time $t$.
Note that for fixed $x$, the set in \eqref{gap:eq:detection} is closed. Indeed, if $t_n\to t$ and $x=(1-\lambda_n)a(t_n)+\lambda_n b(t_n)$, take a subsequence with $\lambda_n\to\lambda\in[0,1]$ and use continuity. Thus $T_{a,b}(x)<\infty$ implies $x\in[a(T_{a,b}(x)),b(T_{a,b}(x))]$.

For a nonempty set of possible targets $F\subseteq\R^2\setminus\{O\}$, write
\begin{equation}\label{gap:eq:restricted}
\CR_F(a,b):=\sup_{x\in F}\frac{T_{a,b}(x)}{\norm x},~
C(F):=\inf_{(a,b)\text{ satisfying }\eqref{gap:eq:admissible}}\CR_F(a,b).
\end{equation}
The paths in \eqref{gap:eq:restricted} may depend on the candidate set $F$, but not on the hidden target's identity. They represent the schedule prescribed by a deterministic algorithm under the hypothetical assumption that no target is ever detected. Actual execution can stop at the first detection.
Let
\begin{equation}\label{gap:eq:circles}
S_r:=\{x\in\R^2:\norm x=r\}\quad(r>0),~
\Cunknown:=C(\R^2\setminus\{O\}).
\end{equation}
In the sequel, we invoke Theorem~\ref{thm:known_tight}, namely
\begin{equation}\label{gap:eq:known}
C(S_r)=\Cknown=C_0:=1+\sqrt3+\frac\pi6
\approx3.255649583<4\qquad(r>0).
\end{equation}
As previously shown, if $ru\in[a(t),b(t)]$, where $\norm u=1$, then $\max\{u\cdot a(t),u\cdot b(t)\}\ge r$; continuity implies that one robot has already touched the tangent line $u\cdot z=r$. 

Two facts will be used repeatedly: scaling space and time preserves the optimal ratio, and restricting the target set cannot increase it. Specifically,
\begin{align}
& \label{gap:eq:scaling} C(\lambda F)=C(F),~
a_\lambda(t)=\lambda a(t/\lambda),~
b_\lambda(t)=\lambda b(t/\lambda), \quad(\lambda>0).\\
& \label{gap:eq:restriction} 
F\subseteq G\quad\Longrightarrow\quad C(F)\le C(G).
\end{align}
Fix the target set
\begin{equation}\label{gap:eq:E}
E:=S_{1/8}\cup S_{1/4}\cup S_1,~ C_E:=C(E).
\end{equation}
The robots may know all three radii $1/8, 1/4, 1$, but are not told which circle contains the target. Now we state the main result which is based on the three-circle construction and differentiates between search with ``known'' and ``unknown'' $d$. 

\begin{theorem}\label{gap:thm:main}
Consider the target set $E$ defined in~\eqref{gap:eq:E}. In the deterministic model \eqref{gap:eq:admissible}--\eqref{gap:eq:restricted},
\begin{equation}\label{gap:eq:strict}
C_0+\delta_0\le C_E\le\Cunknown,
\qquad \delta_0:=\frac{\sqrt{17}}4-1=0.0307764064\ldots.
\end{equation}
Consequently, the constant
\begin{equation}\label{gap:eq:epsilon}
c:=\frac{C_E-C_0}{2}
\end{equation}
is finite and strictly positive, and satisfies
\begin{equation}\label{gap:eq:gap}
\Cknown+c<C_E\le\Cunknown.
\end{equation}
In particular,
\begin{equation}\label{eq:unknown_direct}
\Cunknown\ge\sqrt3+\frac\pi6+\frac{\sqrt{17}}4=3.2864259895\ldots.
\end{equation}
The number $c$ is independent of the algorithm and of the target, and satisfies
\begin{equation}\label{gap:eq:epsilonbounds}
\frac{\sqrt{17}-4}{8}\le c\le0.9875586419\ldots<1.
\end{equation}
\end{theorem}

The rest of this subsection develops the proof of Theorem~\ref{gap:thm:main} and supplementary structural results. We first explain the qualitative obstruction: equality on the outer circle forces one straight initial trajectory (Lemma~\ref{gap:lem:straight}), and four perpendicular targets exclude equality on $E$ (Lemmas~\ref{gap:lem:four}~and~\ref{gap:lem:noequality}). The exact-equality and compactness arguments (Lemmas~\ref{gap:lem:straight},
\ref{gap:lem:noequality}, and~\ref{gap:lem:attainment}) provide supplementary
explanations. The quantitative lower bound itself needs only the four-point
lemma and the narrow-strip lemma.
We then quantify the initial-path constraint using a narrow strip, proving
the stated explicit gap 
(Lemma~\ref{gap:lem:strip}).

\begin{figure}[htbp]
\centering
\begin{tikzpicture}[scale=2.35,>=Latex,font=\small]
\draw[gray!65,thick] (0,0) circle(1);
\draw[orange!80!black,thick] (0,0) circle(.25);
\draw[purple!80!black,thick] (0,0) circle(.125);
\draw[->,gray!65] (0,-1.12)--(0,1.15) node[above] {$v$};
\draw[->,blue!75!black,very thick] (0,0)--(1,0) node[right] {$u=a(1)$};
\node[blue!75!black,fill=white,inner sep=2pt] at (.66,.13) {$a(t)=tu$};
\foreach \y/\lab in {.25/{2rv},.125/{rv},-.125/{-rv},-.25/{-2rv}} {
 \fill[red!75!black] (0,\y) circle(.015);
 \node[anchor=east,fill=white,inner sep=1pt,font=\footnotesize] at (-.12,\y) {$\lab$};
}
\draw[gray!70] (0.707,.707)--(1.06,.83) node[right,black] {$S_1$};
\draw[orange!80!black] (.177,.177)--(.60,.48) node[right,black] {$S_{1/4}$};
\draw[purple!80!black] (.108,-.0625)--(.61,-.43) node[right,black] {$S_{1/8}$};
\draw[gray] (.055,0)--(.055,.055)--(0,.055);
\node[align=center] at (-.72,-.87) {$r=1/8$\\$v\perp u$};
\end{tikzpicture}
\caption{The candidate set $E$ is fixed before the algorithm is chosen. If an algorithm had ratio $C_0$ on $E$, its performance on $S_1$ would force one robot to follow the blue straight segment during $0\le t\le1$. The perpendicular direction $v$ may then be chosen from that trajectory. All four red targets belong to the already fixed set $E$, and their deadlines precede time $1$.}
\label{gap:fig:circles}
\end{figure}
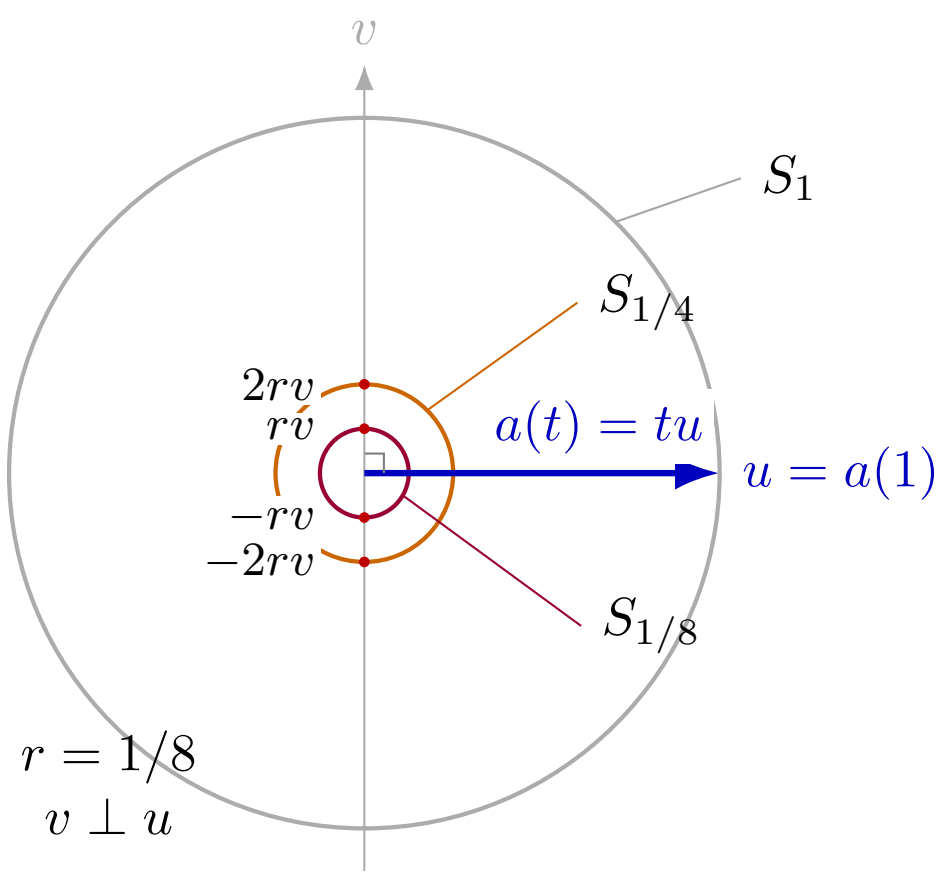

\begin{lemma}\label{gap:lem:straight}
Suppose an admissible pair $a,b$ satisfies
\begin{equation}\label{gap:eq:outerdeadline}
T_{a,b}(x)\le C_0\qquad\text{for every }x\in S_1.
\end{equation}
Then, after possibly exchanging the robot names, there is a unit vector $u$ such that
\begin{equation}\label{gap:eq:straight}
a(t)=tu\qquad(0\le t\le1).
\end{equation}
\end{lemma}
\begin{proof}
Define the first time either endpoint reaches the unit circle by
\begin{equation}\label{gap:eq:tau}
\tau:=\inf\{t\ge0:\max(\norm{a(t)},\norm{b(t)})=1\}.
\end{equation}
This time exists and is finite: detection of any unit-circle point requires at least one endpoint to have norm at least $1$, and continuity forces a preceding endpoint contact with $S_1$. The speed constraints imply $\tau\ge1$. At time $\tau$, both norms are at most $1$, and one is exactly $1$.

Before $\tau$, neither robot has norm $1$, and for every $\lambda\in[0,1]$,
\[
\norm{(1-\lambda)a(t)+\lambda b(t)}
\le\max\{\norm{a(t)},\norm{b(t)}\}<1\qquad(t<\tau).
\]
Thus no point of $S_1$ is detected before $\tau$.

We claim that $\tau =1$. Indeed, suppose on the contrary that $\tau>1$. Replace the initial motions by straight paths to the endpoint positions at time $\tau$, and then replay the original schedule:
\begin{equation}\label{gap:eq:compressed}
\widetilde a(t)=
\begin{cases}
t\,a(\tau),&0\le t\le1,\\
a(t+\tau-1),&t\ge1,
\end{cases}
\qquad
\widetilde b(t)=
\begin{cases}
t\,b(\tau),&0\le t\le1,\\
b(t+\tau-1),&t\ge1.
\end{cases}
\end{equation}
Both paths in \eqref{gap:eq:compressed} satisfy \eqref{gap:eq:admissible}: their initial speeds are at most $1$, the later portions are time translates of admissible paths, and the positions agree at the join $t=1$.
Every original unit-circle detection occurs at some $s\ge\tau$ and is reproduced at time $1+s-\tau$. By \eqref{gap:eq:outerdeadline}, the new pair therefore detects every point of $S_1$ by
\begin{equation}\label{gap:eq:compressiontime}
1+C_0-\tau<C_0.
\end{equation}
This contradicts $C(S_1)=C_0$ from \eqref{gap:eq:known}. Hence $\tau=1$, which proves the claim.

Relabel so that $a(1)=u$ with $\norm u=1$. For $0\le t\le1$, the $1$-Lipschitz property gives
\(
\norm{a(t)}\le t,~ \norm{u-a(t)} = \norm{a(1)-a(t)} \le1-t.
\)
Consequently,
\(
1=\norm u\le\norm{a(t)}+\norm{u-a(t)}\le t+(1-t)=1,
\)
and all inequalities are equalities. For $0<t<1$, equality in the Euclidean triangle inequality places $a(t)$ on the segment $[O,u]$, at distance exactly $t$ from $O$. Thus $a(t)=tu$; the endpoint cases follow as well.
\end{proof}

Note that Lemma~\ref{gap:lem:straight} does not assert uniqueness of the known-radius optimal algorithm. It only establishes one property shared by every pair satisfying \eqref{gap:eq:outerdeadline}. The compressed pair in \eqref{gap:eq:compressed} is used solely as a competitor for searching $S_1$; it need not preserve detections on the smaller circles.

\subsubsection{A four-point obstruction on a line}\label{gap:sec:four}
Next, in Lemma~\ref{gap:lem:four}, we prove a general result concerning four-point line-search that will be used later.
\begin{lemma}[Four-point line-search bound]\label{gap:lem:four}
Let $y:[0,\infty)\to\R$ be $1$-Lipschitz with $y(0)=0$. For $z\in\R$, define
\begin{equation}\label{gap:eq:scalarhit}
\sigma_y(z):=\inf\{t\ge0:y(t)=z\},
\end{equation}
again with $\inf\varnothing=+\infty$. For every $r>0$,
\begin{equation}\label{gap:eq:fourbound}
\max_{z\in\{-2r,-r,r,2r\}}\frac{\sigma_y(z)}{|z|}\ge4.
\end{equation}
\end{lemma}
\begin{proof}
If any required level is never reached, \eqref{gap:eq:fourbound} is immediate. Otherwise all four first hitting times are finite and attained. By replacing $y$ by $-y$ if necessary, assume that $r$ is visited before $-r$. In the rest of the proof we refer to Figure~\ref{gap:fig:four}.
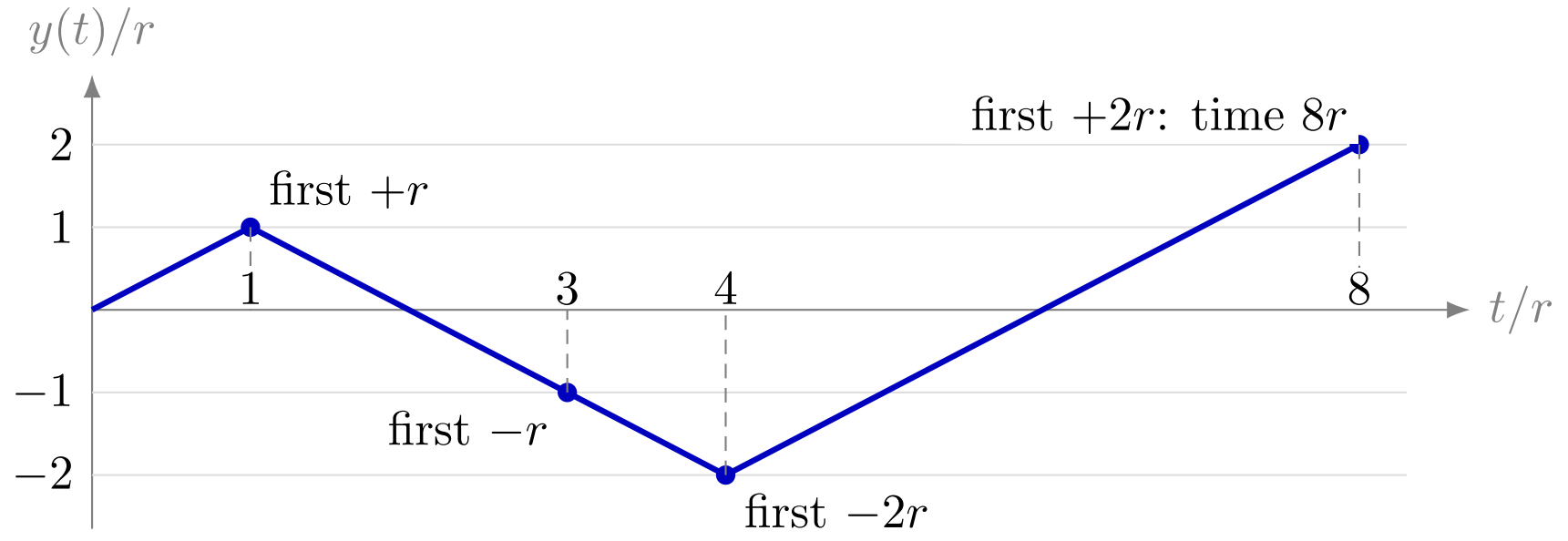
\begin{figure}[htbp]
\centering
\begin{tikzpicture}[x=1.15cm,y=.60cm,>=Latex,font=\small]
\draw[->,gray] (0,0)--(8.7,0) node[right] {$t/r$};
\draw[->,gray] (0,-2.65)--(0,2.85) node[above] {$y(t)/r$};
\foreach \y in {-2,-1,1,2}{\draw[gray!25] (0,\y)--(8.3,\y);\node[left] at (0,\y) {$\y$};}
\draw[blue!75!black,very thick] (0,0)--(1,1)--(3,-1)--(4,-2)--(8,2);
\foreach \x/\y in {1/1,3/-1,4/-2,8/2}{\fill[blue!75!black] (\x,\y) circle(2pt);\draw[densely dashed,gray] (\x,\y)--(\x,0);\node[above,fill=white,inner sep=1pt] at (\x,0) {$\x$};}
\node[above right] at (1,1) {first $+r$};
\node[below left] at (3,-1) {first $-r$};
\node[below right] at (4,-2) {first $-2r$};
\node[above left,fill=white,inner sep=2pt] at (8,2) {first $+2r$: time $8r$};
\end{tikzpicture}
\caption{The shortest itinerary for the better outer-level order in Case 2 of Lemma~\ref{gap:lem:four}. Its successive travel times are $r$, $2r$, $r$, and $4r$. The last target has distance $2r$ and cannot be reached before time $8r$, giving ratio $4$. Detours or waiting can only increase these times.}
\label{gap:fig:four}
\end{figure}

\emph{Case 1: $2r$ is visited before $-r$.} The path must first cover distance at least $2r$ from $0$ to $2r$, and then distance at least $3r$ to $-r$. Hence
\begin{equation}\label{gap:eq:caseone}
\sigma_y(-r)\ge2r+3r=5r,
\end{equation}
which yields a ratio at least $5$.

\emph{Case 2: $-r$ is visited before $2r$.}
Reaching $-r$ after first visiting $r$ requires time at least $r+2r=3r$. Neither outer level has yet been visited: $2r$ has not been visited by the case assumption, and visiting $-2r$ would require crossing $-r$ first.

Starting at $-r$, if the next outer level reached is $-2r$, visiting both outer levels requires at least $r+4r=5r$ more time. If the next outer level is $2r$, the required additional time is at least $3r+4r=7r$. Therefore
\begin{equation}\label{gap:eq:casetwo}
\max\{\sigma_y(-2r),\sigma_y(2r)\}\ge3r+5r=8r.
\end{equation}
The corresponding outer target has distance $2r$, giving ratio at least $4$.
\end{proof}

\subsubsection{Excluding equality on the three circles}\label{gap:sec:exclude}
\begin{lemma}[No pair achieves $C_0$ on $E$]\label{gap:lem:noequality}
There are no admissible paths $a,b$ satisfying
\begin{equation}\label{gap:eq:allE}
T_{a,b}(x)\le C_0\norm x\qquad(x\in E).
\end{equation}
\end{lemma}
\begin{proof}
Suppose, for a contradiction, that Inequality~\eqref{gap:eq:allE} holds. Because $S_1\subset E$, Lemma~\ref{gap:lem:straight} applies. Relabel the robots so that $a(t)=tu$ for $0\le t\le1$, where $\norm u=1$.

Choose a unit vector $v$ perpendicular to $u$, and put $r=1/8$. The four targets
\begin{equation}\label{gap:eq:fourtargets}
rv,\quad-rv,\quad2rv,\quad-2rv
\end{equation}
all belong to $E$. Under \eqref{gap:eq:allE}, their detection deadlines are at most
\begin{equation}\label{gap:eq:early}
2rC_0=\frac{C_0}{4}<1.
\end{equation}
Thus every relevant detection occurs while $v\cdot a(t)=0$.

Define the scalar projection
\begin{equation}\label{gap:eq:projection}
y(t):=v\cdot b(t)\qquad(0\le t\le1).
\end{equation}
This function starts at zero and is $1$-Lipschitz, since
$|v\cdot(b(t)-b(s))|\le\norm{b(t)-b(s)}\le|t-s|$.
Extend it constantly for $t\ge1$, preserving that property.

Let $z\in\{\pm r,\pm2r\}$, and let $t_z=T_{a,b}(zv)<1$. At this detection time there is a $\lambda\in[0,1]$ with
\begin{equation}\label{gap:eq:projectincidence}
zv=(1-\lambda)a(t_z)+\lambda b(t_z),\qquad
z=\lambda y(t_z).
\end{equation}
Here $\lambda>0$, because $z\ne0$. If $z>0$, \eqref{gap:eq:projectincidence} gives $y(t_z)\ge z$; if $z<0$, it gives $y(t_z)\le z$. By continuity from $y(0)=0$, the scalar path has reached level $z$ by time $t_z$. Therefore
\begin{equation}\label{gap:eq:scalardeadlines}
\sigma_y(z)\le T_{a,b}(zv)\le C_0|z|
\qquad(z\in\{\pm r,\pm2r\}).
\end{equation}
Equation~\eqref{gap:eq:scalardeadlines} contradicts Lemma~\ref{gap:lem:four}, because $C_0<4$.
\end{proof}

The direction $v$ in this proof depends on the proposed trajectories, which is legitimate for a worst-case lower bound. The candidate set $E$ itself does not depend on them: it contains the four points \eqref{gap:eq:fourtargets} for every possible choice of $v$. See Figure~\ref{gap:fig:circles}.

\subsubsection{Compactness and the uniform positive gap}\label{gap:sec:compact}

Next we use compactness to show that the infimum is attained for the target sets.
\begin{lemma}\label{gap:lem:attainment}
Let $F\subseteq\R^2\setminus\{O\}$ be nonempty and bounded, and assume $C(F)<\infty$. Then an admissible pair attains the infimum in \eqref{gap:eq:restricted}:
\begin{equation}\label{gap:eq:attainment}
\CR_F(a,b)=C(F).
\end{equation}
\end{lemma}
\begin{proof}
Put $c_F=C(F)$ and choose $R>0$ with $\norm x\le R$ for all $x\in F$. By the definition of infimum, choose admissible pairs $a_n,b_n$ such that
\begin{equation}\label{gap:eq:minseq}
T_{a_n,b_n}(x)\le(c_F+1/n)\norm x\qquad(x\in F).
\end{equation}
Set $H=(c_F+1)R$. All deadlines in \eqref{gap:eq:minseq} lie in $[0,H]$. The paths on this interval are uniformly bounded, because $\norm{a_n(t)},\norm{b_n(t)}\le t\le H$, and uniformly equicontinuous by \eqref{gap:eq:admissible}. Apply Arzel\`a--Ascoli to the paired maps $(a_n,b_n):[0,H]\to\R^4$. After passage to a subsequence, both paths converge uniformly to $a,b$. The limit paths start at $O$ and retain the $1$-Lipschitz bounds.

Fix any $x\in F$. For each index of this subsequence choose a detection time $s_n$ and segment parameter $\lambda_n$ with
\begin{equation}\label{gap:eq:witnesses}
0\le s_n\le(c_F+1/n)\norm x,\quad 0\le\lambda_n\le1,
\qquad x=(1-\lambda_n)a_n(s_n)+\lambda_n b_n(s_n).
\end{equation}
Such witnesses exist by attainment of finite detection times. Compactness of $[0,H]\times[0,1]$ gives a further subsequence on which $s_n\to s$ and $\lambda_n\to\lambda$. Then $s\le c_F\norm x$. Moreover,
\[
\norm{a_n(s_n)-a(s)}
\le\sup_{t\in[0,H]}\norm{a_n(t)-a(t)}+|s_n-s|\longrightarrow0,
\]
and the same holds for $b_n$. Passing to the limit in \eqref{gap:eq:witnesses} gives
\begin{equation}\label{gap:eq:limitwitness}
x=(1-\lambda)a(s)+\lambda b(s),\qquad s\le c_F\norm x.
\end{equation}

The limiting paths $a,b$ were fixed before $x$ was chosen. Only the subsequence of witnesses may depend on $x$; thus \eqref{gap:eq:limitwitness} holds for every $x\in F$ for this same pair. There is no need to select one subsequence of witnesses simultaneously for an uncountable set of targets.

Extend both limiting paths constantly after time $H$. This preserves \eqref{gap:eq:admissible} and every detection in \eqref{gap:eq:limitwitness}. Their ratio on $F$ is at most $c_F$, and cannot be smaller than its infimum $C(F)=c_F$. This proves \eqref{gap:eq:attainment}.
\end{proof}

\subsubsection{Quantitative initial-path estimate}
\begin{lemma}[A narrow strip before first unit-circle contact]
\label{gap:lem:strip}
Suppose an admissible pair discovers every point of $S_1$ by time $C$.
Let $\tau$ be the first time either endpoint reaches $S_1$. After relabeling,
write $a(\tau)=u$ with $\norm u=1$. Then
\begin{equation}\label{gap:eq:taubound}
1\le\tau\le1+C-C_0.
\end{equation}
For every unit vector $v\perp u$ and every $0\le t\le\tau$,
\begin{equation}\label{gap:eq:strip}
|v\cdot a(t)|\le\frac{\sqrt{\tau^2-1}}2.
\end{equation}
\end{lemma}
\begin{proof}
No unit-circle target is detected before $\tau$. The compressed paths in
\eqref{gap:eq:compressed} therefore discover all of $S_1$ by time
$1+C-\tau$. The known-radius lower bound implies
$C_0\le1+C-\tau$, thus proving \eqref{gap:eq:taubound}.
Write $a(t)=xu+yv$. Unit speed gives
\[
\sqrt{x^2+y^2}+\sqrt{(1-x)^2+y^2}
=\norm{a(t)}+\norm{u-a(t)}\le t+(\tau-t)=\tau.
\]
Applying the triangle inequality to $(x,y)$ and $(1-x,y)$ yields
$\sqrt{1+4y^2}\le\tau$, which proves \eqref{gap:eq:strip}.
\end{proof}

Finally, we are in a position to prove the main theorem~\ref{gap:thm:main}.
\begin{proof}[Proof of Theorem~\ref{gap:thm:main}]
Recall the parameters $\delta_0 =\frac{\sqrt{17}}4-1, ~c = \frac{C_E-C_0}2, C_0 = 1+\sqrt{3}+\frac{\pi}6$ in the statement of the theorem. Suppose, for a contradiction, that a pair has ratio $C<C_0+\delta_0$ on $E$; necessarily $C\ge C_0$.
Its unit-circle targets are discovered by time $C$, so
Lemma~\ref{gap:lem:strip} supplies an endpoint $a(\tau)=u$ and gives
\[
1\le\tau\le1+C-C_0<1+\delta_0=\frac{\sqrt{17}}4.
\]
Choose a unit vector $v\perp u$ and put $r=1/8$. For $0\le t\le\tau$,
\begin{equation}\label{gap:eq:strictstrip}
|v\cdot a(t)|\le\frac{\sqrt{\tau^2-1}}2<\frac18=r.
\end{equation}
All four targets $zv$, $z\in\{\pm r,\pm2r\}$, belong to $E$ and must
be detected by time $2rC=C/4<1\le\tau$, since $C_0+\delta_0<4$.
At a detection time $t_z$, take projections of the segment incidence:
\[
z=(1-\lambda)v\cdot a(t_z)+\lambda v\cdot b(t_z),
\qquad 0\le\lambda\le1.
\]
If $z>0$, \eqref{gap:eq:strictstrip} implies $v\cdot a(t_z)<r\le z$,
so $v\cdot b(t_z)\ge z$. If $z<0$, it implies
$v\cdot a(t_z)>-r\ge z$, so $v\cdot b(t_z)\le z$.
The $1$-Lipschitz scalar path $y_b(t)=v\cdot b(t)$ starts at zero;
continuity therefore gives $\sigma_{y_b}(z)\le t_z\le C|z|$ for all
four levels. Lemma~\ref{gap:lem:four} forces $C\ge4$, a contradiction.

Thus no admissible pair can have ratio less than $C_0+\delta_0$ on $E$,
and taking the infimum gives $C_E\ge C_0+\delta_0$. Restriction gives
$C_E\le\Cunknown$. Proposition~\ref{rem:constant} below proves that
$C_E<\infty$ and establishes the upper bound in \eqref{gap:eq:epsilonbounds}. Hence
\[
c=\frac{C_E-C_0}{2}\ge\frac{\delta_0}{2}
=\frac{\sqrt{17}-4}{8}=0.0153882032\ldots 
\]
Since $C_0+c=(C_0+C_E)/2<C_E$, this concludes the proof of Theorem~\ref{gap:thm:main}.
\end{proof}

\begin{proposition}[Upper bound on the gap constant]\label{rem:constant}
The constant $c$ in \eqref{gap:eq:epsilon} is finite and satisfies
$c\le0.9875586419309\ldots<1$.
\end{proposition}
\begin{proof}
Let
\[
L(r,R)=\sqrt{(R/\sqrt3-r)^2+(R+r)^2}.
\]
Search radii $1/8,1/4,1$ in order using the known-radius algorithm.
After finishing radius $r$ at $(\pm r,-r)$, move directly to
$(\pm R/\sqrt3,R)$ and run the last three phases for radius $R$.
Those phases cover all of $S_R$, so the deployment phase need not be
repeated. The completion times satisfy
\begin{align*}
t_1&=C_0/8,\\
t_2&=t_1+L(1/8,1/4)+(C_0-2/\sqrt3)/4,\\
t_3&=t_2+L(1/4,1)+C_0-2/\sqrt3.
\end{align*}
Then $C_E\le\max\{8t_1,4t_2,t_3\}=4t_2$, yielding the exact upper bound
\begin{equation}\label{gap:eq:epsilonupper}
c\le \frac{C_0}{4}+2L(1/8,1/4)-\frac1{\sqrt3}
=0.9875586419309\ldots<1.
\end{equation}
For a simple conservative check of the maximum and the final inequality,
use $3.2556<C_0<3.2557$, $0.37549<L(1/8,1/4)<0.37551$, and
$1.29215<L(1/4,1)<1.29216$, which imply
$5.2306<4t_2<5.2309$ and $t_3<4.7010$.
This bounds the
\emph{chosen} $c$, not the full difference between the two optimal ratios.

\end{proof}

\begin{corollary}\label{gap:cor:uniform}
For the positive constant $c$ in \eqref{gap:eq:epsilon}, every deterministic universal search algorithm has some target $x\in E$ such that
\begin{equation}\label{gap:eq:uniform}
T_{a,b}(x)>(C_0+c)\norm x.
\end{equation}
The same assertion holds with $E$ replaced by $S_1\cup S_2\cup S_8$.
\end{corollary}
\begin{proof}
For every pair, its supremum in \eqref{gap:eq:restricted} is at least $C_E>C_0+c$. If every target satisfied the reverse non-strict inequality to \eqref{gap:eq:uniform}, that supremum would be at most $C_0+c$, a contradiction. The final statement follows from \eqref{gap:eq:scaling}, since $8E=S_1\cup S_2\cup S_8$.
\end{proof}

\section{Search with \texorpdfstring{$n\ge3$}{n >= 3} Robots}\label{app:nrobots}
This section extends instantaneous segment discovery to $n\ge3$ robots.
We give the expanding regular-polygon algorithm, compute its exact
competitive ratio, and prove optimality for arbitrary deterministic
trajectories by reduction to shoreline search. We consider both known and unknown target distance.

\subsection{Model and statement of the result}
Robots $0,\ldots,n-1$ start at $O$ and have $1$-Lipschitz trajectories
$p_i:[0,\infty)\to\R^2$. In the base model, every pairwise segment
$[p_i(t),p_j(t)]$ is visible. The polygon interior is not, in general,
instantaneously visible. Our upper-bound algorithm needs only the
segments joining cyclic neighbors $i-1,i,i+1$, with indices modulo $n$.
We assume instantaneous notification of detection to stop the robots;
the objective is discovery, and the Boolean sensor need not localize the target.

For a deterministic algorithm $\mathcal B$, let $V_{\mathcal B}(t)$ be
the union of its visible segments and put
\begin{equation}\label{eq:nrobotCR}
T_{\mathcal B}(x)=\inf\{t\ge0:x\in V_{\mathcal B}(t)\},\qquad
\CR_{\mathcal B}^{(n)}=\sup_{x\ne O}\frac{T_{\mathcal B}(x)}{\norm x},
\qquad \CR_n^*=\inf_{\mathcal B}\CR_{\mathcal B}^{(n)}.
\end{equation}
The infimum is over admissible deterministic algorithms. 
The offline optimum is $\norm x$:
no endpoint is farther than $t$ at time $t$, so neither is any point on
a visible segment, while sending two robots straight to $x$ attains
that time. The lower bound is proved for all-pairs visibility, while the upper
bound uses only cyclic-neighbor segments, so the optimum is the same in both models.

For a supplied $d>0$, define
\begin{equation}\label{eq:nrobotknown}
\CR^*_{n,\mathrm{known}\,d}:=\inf_{\mathcal B_d}\sup_{\|x\|=d}
\frac{T_{\mathcal B_d}(x)}d,\qquad
\CR^*_{n,\mathrm{unknown}\,d}:=\CR_n^*,
\end{equation}
where $\mathcal B_d$ may use $d$ but not the target direction.

\begin{theorem}
\label{thm:nrobots}
For every integer $n\ge3$, the algorithm $\mathcal A_n$ in
Algorithm~\ref{alg:nrobots} does not use the
target distance and has exact worst-case ratio
\begin{equation}\label{eq:nrobotopt}
\CR_{\mathcal A_n}^{(n)}=\CR^*_{n,\mathrm{unknown}\,d}
=\CR^*_{n,\mathrm{known}\,d}=\sec(\pi/n).
\end{equation}
Optimality is among all admissible deterministic trajectories, with
arbitrary turns and speeds at most one. The same optimum holds with only cyclic-neighbor visibility.
\end{theorem}

\subsection{The algorithm and its geometric analysis}
Define
\begin{equation}\label{eq:nrobotdirections}
\beta=\frac\pi n,\qquad
\theta_i=\frac{2\pi i}{n},\qquad
u_i=(\cos\theta_i,\sin\theta_i),\qquad p_i(t)=t u_i.
\end{equation}
At time $t>0$ the robots form a regular $n$-gon of circumradius $t$.
Its edges are the segments monitored by the algorithm, and its inradius is $t\cos\beta$.

\begin{algorithm}[H]
\caption{Expanding regular polygon $\mathcal A_n$}
\label{alg:nrobots}
\begin{algorithmic}[1]
\Require $n\ge3$ robots at $O$, common clock and agreed directions;
\Ensure Discovery of every stationary target $x\ne O$, without knowing $\norm x$
\State Assign robot $i$ the vector $u_i=(\cos(2\pi i/n),\sin(2\pi i/n))$, for $0\le i<n$
\State Assign robot $i$ its two neighbors $i-1$ and $i+1$ modulo $n$
\State Start all robots simultaneously at time $0$
\While{no target has been detected}
\State All robots move continuously and simultaneously: $p_i(t)=t u_i$
\State Robot $i$ continuously monitors $[p_i(t),p_{i-1}(t)]$ and $[p_i(t),p_{i+1}(t)]$
\EndWhile
\State Detecting robots announce detection; all robots stop
\end{algorithmic}
\end{algorithm}

\begin{lemma}[Exact discovery time]\label{lem:nrobotupper}
Let $x$ have polar radius $d>0$ and angle $\theta$. Choose a sector
$[\theta_i,\theta_{i+1}]$ containing its direction, interpreting the last
sector modulo $2\pi$, and set $\varphi=\theta-(\theta_i+\beta)$ in that
sector. Then $|\varphi|\le\beta$ and
\begin{equation}\label{eq:nrobottime}
T_{\mathcal A_n}(x)=\frac{d\cos\varphi}{\cos\beta}.
\end{equation}
Consequently the worst-case ratio is $1/\cos\beta$, attained along
every sector bisector.
\end{lemma}
\begin{proof}
Let $m_i$ be the unit vector along the sector bisector and $w_i$ its
counterclockwise perpendicular. In these coordinates the two endpoints
are $(t\cos\beta,-t\sin\beta)$ and $(t\cos\beta,t\sin\beta)$, and
$x=(d\cos\varphi,d\sin\varphi)$ (see Figure~\ref{fig:nrobots}). Membership in their segment requires
\(
d\cos\varphi=t\cos\beta,~ |d\sin\varphi|\le t\sin\beta.
\)
The first equality gives \eqref{eq:nrobottime}. At that time the second
inequality follows from $|\tan\varphi|\le\tan\beta$, since
$0<\beta\le\pi/3<\pi/2$. 
\begin{figure}[H]
\centering
\begin{tikzpicture}[scale=1.65,>=Latex,font=\small]
\begin{scope}[xshift=-2.6cm]
\foreach \i in {0,...,4}{
 \pgfmathsetmacro{\ang}{18+72*\i}
 \coordinate (P\i) at (\ang:1.65);
 \draw[->,gray!65] (0,0)--(\ang:1.94);
 \fill[blue!70!black] (P\i) circle(1.8pt);
}
\fill[blue!4] (P0)--(P1)--(P2)--(P3)--(P4)--cycle;
\draw[blue!75!black,thick] (P0)--(P1)--(P2)--(P3)--(P4)--cycle;
\foreach \i in {0,...,4}{\fill[blue!70!black] (P\i) circle(1.8pt);}
\draw[gray,dashed] (0,0) circle(1.335);
\fill (0,0) circle(1.5pt) node[below] {$O$};
\node at (0,-2.12) {Expanding polygon ($n=5$)};
\end{scope}
\begin{scope}[xshift=2.2cm]
\coordinate (O) at (0,0);
\coordinate (A) at (-1.2,1.65);
\coordinate (B) at (1.2,1.65);
\coordinate (M) at (0,1.65);
\coordinate (X) at (.65,1.65);
\draw[->,gray!65] (O)--(-1.46,2.01);
\draw[->,gray!65] (O)--(1.46,2.01);
\draw[blue!75!black,very thick] (A)--(B);
\draw[thick] (O)--(A) node[midway,left] {$t$};
\draw[thick] (O)--(B) node[midway,right] {$t$};
\draw[dashed] (O)--(M);
\draw[orange!85!black,thick] (O)--(X) node[pos=.67,right] {$d$};
\draw (-.13,1.65)--(-.13,1.52)--(0,1.52);
\draw (54:.44) arc[start angle=54,end angle=126,radius=.44];
\node[fill=white,inner sep=1pt] at (0,.57) {$2\beta$};
\draw (68.5:.94) arc[start angle=68.5,end angle=90,radius=.94];
\node at (.18,1.08) {$|\varphi|$};
\fill (O) circle(1.5pt) node[below] {$O$};
\fill[blue!75!black] (A) circle(1.8pt) node[above left] {$p_i(t)$};
\fill[blue!75!black] (B) circle(1.8pt) node[above right] {$p_{i+1}(t)$};
\fill (M) circle(1.2pt) node[above] {$M$};
\fill[red!75!black] (X) circle(2pt) node[above] {$x$};
\node at (0,-.52) {$OM=t\cos\beta$};
\node at (0,-2.12) {One sector at detection};
\end{scope}
\end{tikzpicture}
\caption{Only the polygon edges are visible at a given instant. Their
expansion sweeps the plane. In one sector, the projection of the target
onto the sector bisector is $d\cos\varphi=t\cos\beta$. The worst direction
is the bisector, where $x=M$ and $t/d=1/\cos\beta$.}
\label{fig:nrobots}
\end{figure}
Thus detection occurs at the asserted time.
For $t>0$ each polygon edge lies in its corresponding closed angular
sector. A target in the interior of the chosen sector cannot lie on any
other edge. At a shared boundary direction, detection is at the common
vertex at time $d$, also given by \eqref{eq:nrobottime}. This proves that
there is no earlier edge detection. Every other pairwise segment lies
inside the polygon, so it also cannot detect $x$ before the expanding
polygon boundary reaches $x$. Finally $\cos\varphi\le1$, with
equality on the bisector, proving both the upper bound and its attainment.
The trajectories have speed exactly one. 
\end{proof}

\subsection{The unrestricted lower bound}
We use the following shoreline-search bounds as external inputs;
the reduction to segment discovery is proved below.
\begin{theorem}[Shoreline Search~\cite{acharjee2020,dobrev2020}]
\label{thm:nrobotshore}
For deterministic search for an unknown line by $n$ unit-speed robots
starting at $O$, with neither its direction nor its distance supplied,
the competitive ratio is at least $1/\cos(\pi/n)$ for every $n\ge3$.
\end{theorem}

For $n\ge4$ this is the result of~\cite{acharjee2020}. For $n=3$ the bound $2$ is due to~\cite{dobrev2020}; their three-robot bound
holds even when the distance is known.
For a fixed trajectory tuple $\mathbf p=(p_0,\ldots,p_{n-1})$, define
\begin{equation}\label{eq:nrobotH}
H_{\mathbf p}(u;d)=\inf\{t\ge0:\text{some }i\text{ satisfies }u\cdot p_i(t)=d\},
\qquad \norm u=1,\quad d>0.
\end{equation}
This is the first endpoint contact with $\ell_{u,d}=\{z:u\cdot z=d\}$.
Also let $T^{\rm all}_{\mathbf p}(x)$ be the discovery time when every
pairwise segment is visible.
\begin{lemma}[Reduction for $n$ endpoints]\label{lem:nrobotreduction}
For every admissible tuple and every $d>0$, $\norm u=1$,
\begin{equation}\label{eq:nrobotreduction}
H_{\mathbf p}(u;d)\le T^{\rm all}_{\mathbf p}(du)
\le T_{\mathcal B}(du),
\end{equation}
where $\mathcal B$ uses these paths before discovery and monitors any
subset of the pairwise segments.
\end{lemma}
\begin{proof}
If $du\in[p_i(t),p_j(t)]$, some $\lambda\in[0,1]$ satisfies
\(
d=(1-\lambda)u\cdot p_i(t)+\lambda u\cdot p_j(t).
\)
Thus at least one endpoint has projection at least $d$. Its projection
starts at zero and is continuous, so it has reached $d$ by time $t$.
This proves $H_{\mathbf p}(u;d)\le t$ for every pairwise detection time;
taking their infimum proves the first inequality, also when that infimum
is infinite. The second follows because adding visible links cannot delay
detection. In particular, no assertion that the visible set contains the
whole convex hull is needed.
\end{proof}

\begin{lemma}[Known-radius lower bound for $n\ge4$]\label{lem:nknown}
For $n\ge4$ robots, every deterministic segment-search algorithm,
even one supplied with $d>0$, has worst-case detection time at least
$d/\cos(\pi/n)$ on $S_d$.
\end{lemma}
\begin{proof}
Fix $d>0$, put $\beta=\pi/n\le\pi/4$, and let $0<T<d/\cos\beta$.
For any $n$ prescribed trajectories, choose an open angular cone of width
$2\beta$ containing none of the robot positions at time $T$. Such a cone
exists because the circular gaps between at most $n$ nonzero endpoint
directions sum to $2\pi$; endpoints at $O$ do not lie in an open cone.
Let $u$ bisect that cone, and consider the line $\ell=\{z:u\cdot z=d\}$.

A contact point on $\ell$ outside the open cone has norm at least
$d/\cos\beta>T$, so cannot be reached by time $T$. If a robot touches
$\ell$ inside the cone, it must subsequently reach the cone boundary by
$T$, because its final position is outside the interior. In coordinates
along $u$ and its perpendicular, write the contact as $K=(d,y)$, where
$|y|<d\tan\beta$. The distance from $K$ to the closer boundary ray is
$d\sin\beta-|y|\cos\beta$; the orthogonal projection lies on that ray.
The robot's traveled length up to its exit is consequently at least
\[
f(|y|):=\sqrt{d^2+y^2}+d\sin\beta-|y|\cos\beta.
\]
For $0\le z\le d\tan\beta$,
\[
f'(z)=\frac z{\sqrt{d^2+z^2}}-\cos\beta
\le\sin\beta-\cos\beta\le0.
\]
Hence $f(|y|)\ge f(d\tan\beta)=d/\cos\beta>T$, again impossible.
No robot has touched this line by $T$. The endpoint-contact reduction
then implies that the tangency target $du$ has not been segment-detected
by $T$. Since this holds for every $T<d/\cos\beta$, the worst-case time
on $S_d$ is at least $d/\cos\beta$, even for paths chosen using $d$.
 This argument adapts the geometric method in Section 2 of~\cite{acharjee2020}
to a fixed target radius.

\end{proof}

\begin{proof}[Proof of Theorem~\ref{thm:nrobots}]
Lemma~\ref{lem:nrobotupper} proves the upper bound. For the lower bound, fix any deterministic algorithm $\mathcal B$
for the unknown-distance segment problem. Let $\mathbf p$ be the paths
it prescribes under the hypothetical assumption that no target is ever
detected. This schedule is independent of the unknown target location.
Construct a shoreline-search strategy by following exactly these paths
until an endpoint reaches the actual hidden line. Do not stop or change
the schedule at hypothetical point-target detections. This is an
admissible shoreline strategy that uses neither the unknown line's
distance nor its direction.

For every $u,d$, Lemma~\ref{lem:nrobotreduction} compares the line-contact
time with the segment discovery time for its tangency target $du$.
Dividing by $d$ and taking suprema yields
\begin{equation}\label{eq:nrobottransfer}
\sup_{\norm u=1,\,d>0}\frac{H_{\mathbf p}(u;d)}d
\le\sup_{x\ne O}\frac{T_{\mathcal B}(x)}{\norm x}
=\CR_{\mathcal B}^{(n)}.
\end{equation}
Theorem~\ref{thm:nrobotshore} bounds the left side below by
$1/\cos(\pi/n)$. Since $\mathcal B$ was arbitrary, the same bound holds
after taking the infimum over segment-search algorithms. This matches
the upper bound for unknown distance. For known distance and $n=3$,
apply the same tangent-line reduction at that fixed $d$ to the
known-distance shoreline bound $2$ from~\cite{dobrev2020}. For $n\ge4$,
Lemma~\ref{lem:nknown} gives the matching fixed-distance lower bound.
The polygon algorithm does not use $d$, so supplies the upper bound in
both cases. This proves \eqref{eq:nrobotopt}. The lower bounds allow every pairwise segment to be visible, whereas
the algorithm requires only cyclic-neighbor segments. This proves the
assertion for both visibility conventions.
\end{proof}

\section{Conclusion}\label{sec:conclusion}
We have studied deterministic discovery by $n\ge2$ unit-speed robots in the plane, where a target is detected when it lies on an instantaneous segment joining a pair of robots. For $n=2$, the knowledge of the target radius $d$ leads to two fundamentally different outcomes. When $d$ is known, the symmetric four-phase construction attains the exact optimum $\CR^*_{\mathrm{known}\,d}=1+\sqrt3+\pi/6$. When $d$ is unknown, a search strategy must cover all radii without using $d$ in its motion schedule. The three-circle obstruction and quantitative narrow-strip estimate give a strictly stronger lower bound than the known-radius optimum. We obtain
\(
\sqrt3+\frac\pi6+\frac{\sqrt{17}}4\;\le\;\CR^*_{\mathrm{unknown}\,d}
\;\le\;8.928516954\ldots.
\)
For the latter model we analyzed a scale-invariant construction using two
antipodal logarithmic spirals. It attains the displayed upper bound, which
is optimal \emph{within that logarithmic-spiral family}. 
For $n\ge3$, the expanding regular-polygon algorithm attains the exact
optimum $\sec(\pi/n)$ for both known and unknown target radius, even
using only cyclic-neighbor segments.
Determining the unknown-radius optimum for $n=2$ robots remains the principal open problem.
Subsection~\ref{subsec:unknown_lower} gives an explicit separation of at
least $(\sqrt{17}-4)/4$ between the two optimal ratios. Improving the
choice of radii or strengthening the projection obstruction may yield
larger quantitative lower bounds. A complementary direction is to improve
upper bounds by allowing nonsymmetric, nonantipodal, or nonlogarithmic
trajectories. Any matching lower-bound proof must account for the extra
coverage provided by a segment with \emph{both} endpoints moving; a result
restricted to diameters or supporting-line contact would not by itself
resolve the unrestricted problem.

\section*{Acknowledgements}
Many thanks to K. Georgiou for useful conversations in the initial stages of this research.

\end{document}